# Suprafroth (Superconducting Froth)

Ruslan Prozorov[1,2], Andrew F. Fidler[1,3], Jacob Hoberg[1,2], and Paul C. Canfield[1,2]

30 November 2007

[1]*Ames Laboratory, Ames, IA 50011*
[2]*Department of Physics & Astronomy, Iowa State University, Ames, IA 50011*
[3]*Department of Physics, Albion College, Albion, Michigan, 59224*

**The structure and dynamics of froths have been subjects of intense interest due to the desire to understand the behaviour of complex systems where topological intricacy prohibits exact evaluation of the ground state. The dynamics of a traditional froth involves drainage and drying in the cell boundaries, thus it is irreversible. We report a new member to the froths family: suprafroth, in which the cell boundaries are superconducting and the cell interior is normal phase. Despite very different microscopic origin, topological analysis of the structure of the suprafroth shows that statistical von Neumann's and Lewis' laws apply. Furthermore, for the first time in the analysis of froths there is a global measurable property, the magnetic moment, which can be directly related to the suprafroth structure. We propose that this suprafroth is a new, model system for the analysis of the complex physics of two-dimensional froths – with magnetic field and temperature as external (reversible) control parameters.**

The fundamental physics and chemistry of complex systems along with practical issues related to the stability of froths (e.g., insulating foams, ultra-light alloys, fire extinguishers etc) have made them a topic of broad interest[1,2]. Though precise microscopic analysis of froth dynamics is difficult, general laws, that take into account both the topological constrains and physics and chemistry of the froth matter, have been developed. This gives rise to the hope that, whereas the behaviour of individual cells may be unpredictable, the overall system can be described by relatively simple rules. Various froths have been studied, and in most cases the coarsening parameter has been time and the microscopic mechanism has been diffusion of vapour molecules between the cells as well as drainage of the liquid from the cell walls. In the case of a magnetic froth, however, the coarsening is promoted by the applied magnetic field. Two types of magnetic froths have been known: (i) a ferrofluid, in an immiscible liquid, stimulated by an applied alternating magnetic field[2,3] and (ii) the magnetic domain structure observed in transparent ferromagnetic garnets[1,4]. However, in both cases, only limited analysis exploring the similarities between time-dependent coarsening of the conventional froth and magnetic field - induced coarsening in the magnetic froth was done.

Here we used low temperature, magneto-optical imaging of superconducting lead to add an entirely new type of magnetic froth, formed in clean, type-I superconductors. Specifically, in this froth the cell boundaries consist of superconducting phase whereas cell interiors are normal state metal filled with magnetic flux. Unlike any froth studied before, this superconducting froth involves only electrons – normal and paired in Cooper pairs and a magnetic field. We propose to abbreviate this new topological phase as "suprafroth" in the spirit of early discoveries when superconductors were called "supraconductors". The coarsening of the suprafroth is promoted either by increasing applied magnetic field or increasing temperature and does not involve mass transport, thus allows for the study of the topological hysteresis decoupled from the hysteresis of the "aging" of the froth matter. In addition, it is important to note that, unlike all previous cases, here we know macroscopic laws governing the behavior of clean superconductors. Thus we are able to check the compatibility of the statistical laws of cellular evolution with the macroscopic behavior as function of magnetic field and temperature.

Lead is a type-I superconductor in which the interface energy between the normal and superconducting phases is positive, so the system wants to avoid or minimize the number of such boundaries. On the other hand, due to the non-zero demagnetization factor, $N$, the magnetic field on the edge of a disc-shaped lead crystal, $H_e$, is larger than the applied field by $H_e = H/(1-N)$. When $H_e$ exceeds the critical field, $H_c$, magnetic flux penetrates the specimen and an inhomogeneous, intermediate state is formed. It has recently been shown that in pinning-free bulk type-I superconductors the intermediate state in a consists of tubes[5,6] rather than a textbook, laminar structure[7]. At high enough densities the tubes evolve into a well-defined superconducting froth as shown in Figure 1. It should be noted that tubular structure was observed and studied before[6,8,9], but mostly in films[8] where they do not evolve into the suprafroth, most likely because of the low values of the magnetic Bond number (0.3-350) and pinning. In our case of clean bulk crystals, the magnetic Bond number is >2800, which means that dipolar interaction between normal domains can be neglected and tubes interact in the interior repelling each other by super-currents flowing around them enabling formation of the suprafroth. It will be interesting to systematically study the evolution of the suprafroth with sample thickness.

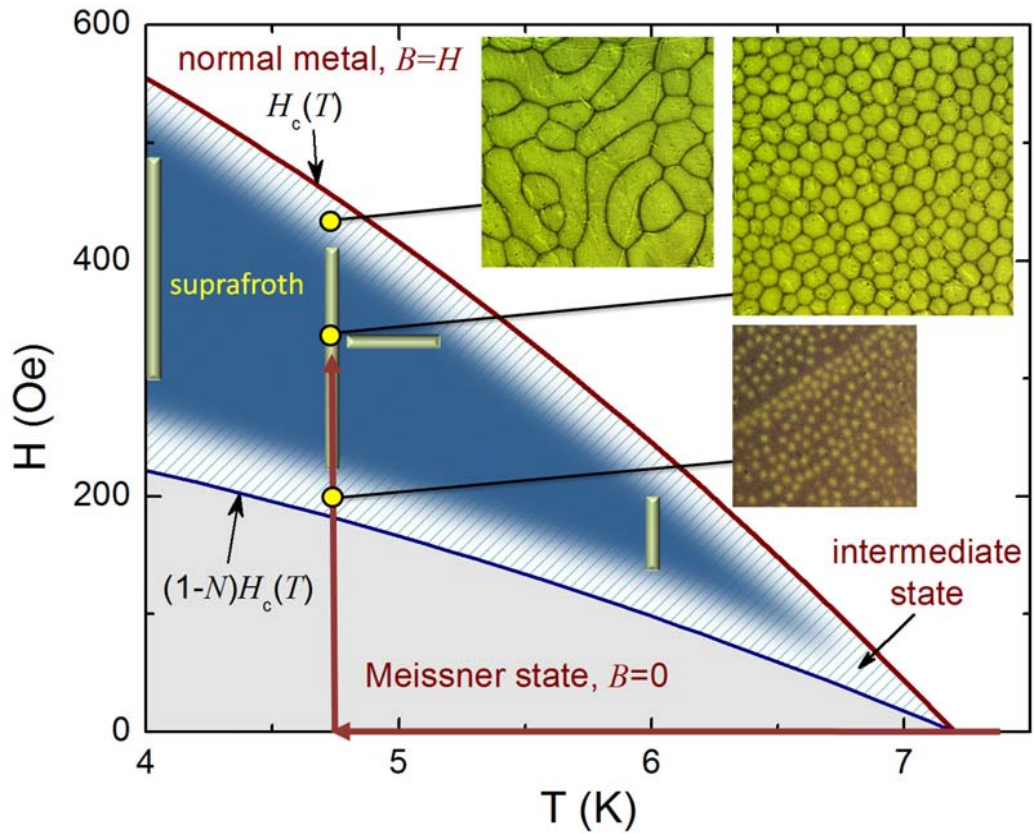

Figure 1. Phase diagram of a superconducting lead disk. Below the $(1-N)H_c(T)$ line, there is a genuine superconducting (Meissner) state in which no magnetic flux exists. Above the $H_c(T)$ line, there is no superconductivity. The shaded region between these lines is the intermediate state. Solid color inside this region shows where the suprafroth is stable. There is no well-defined boundary for the suprafroth region,  - at small magnetic fields, separate tubes do not yet form a polygonal structure (lowest insert), whereas at higher fields, the structure evolves into more rounded and elongated objects that are difficult to call "froth" (upper left insert). Magneto-optical images illustrate this pattern evolution for magnetic field increasing after cooling in zero field (4.8 K path shown). Vertical lines at 4.0, 4.8 and 6.0 K as well as the horizontal line at 320 Oe show where images and data used in this work were acquired.

The magneto-optical images show a clear pattern evolution upon increase of the applied magnetic field. Initially, separate flux tubes are injected into the sample (Figure 1, lower inset). As their number increases, they coalesce and grow. As the field grows, the repulsive forces between the tubes ultimately lead to formation of the suprafroth with well-defined polygonal cellular structure (Figure 1, large inset). For magnetic fields that approach the upper limit of superconductivity, $H_c$, this structure ultimately degrades and forms extended concave cells with rounded boundaries. The suprafroth structure is very different from that seen for the more common bulk, type-II, superconductors where interface energy is negative and magnetic field inside the material exists in form of Abrikosov vortices each carries a single flux quanta and no suprafroth is possible.

The coarsening of a suprafroth with applied magnetic field is shown in Figure 2 (to row). The images were obtained at $T = 6$ K when, after zero-field cooling, the indicated magnetic fields were applied. The bottom row of Figure 2 shows coarsening with increasing temperature sampled at $H = 320$ Oe. The observed patterns were traced and converted into black and white images as shown in Figure 3. Traced images were analyzed in terms of cell statistics. Our typical region of interest was 2x2 mm and contained up to 160 cells.

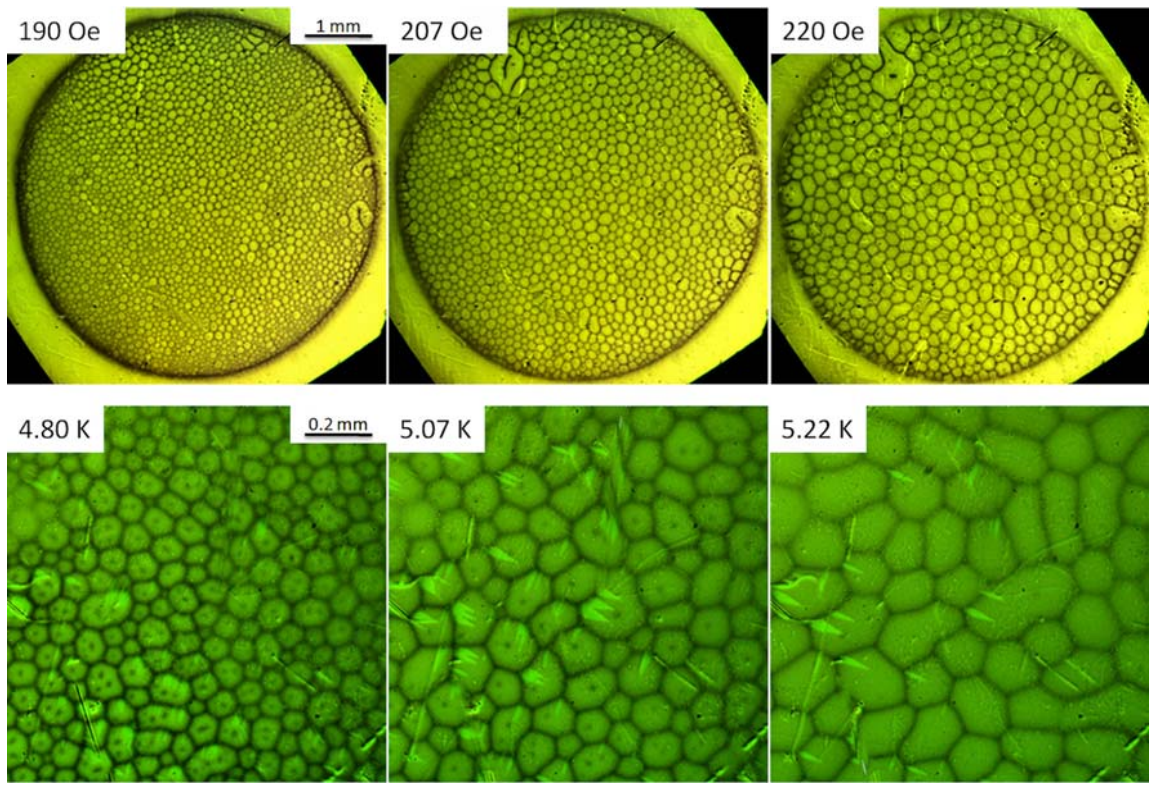

Figure 2. (Top row) Coarsening of the suprafroth with increasing magnetic field at $T=6.0$ K showing the entire sample (5 mm disc). (Bottom row) coarsening with increasing temperature at $H=320$ Oe shown in a 1x1 mm$^2$ region of the disc.

Before we proceed to the discussion of the statistical laws, though, it is important to examine the question of reversibility. In conventional froths, the parameter that is varied to study coarsening is time. The physical processes of drainage and diffusion responsible for coarsening are inherently irreversible.

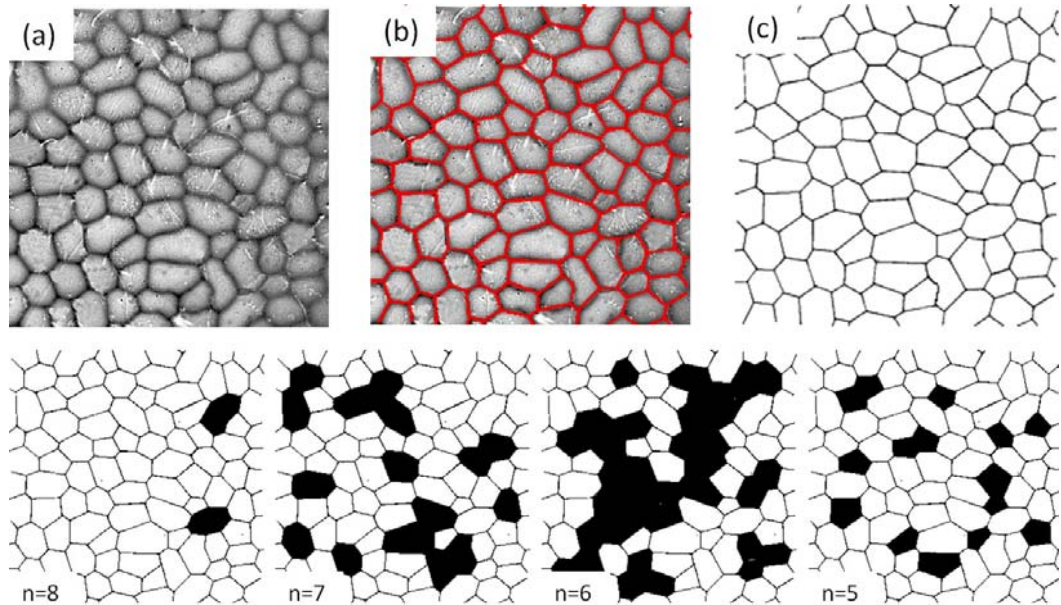

Figure 3. 2x2 mm$^2$ image measured at 4.8 K and 390 Oe. (a) original image; (b) superimposed trace of the boundaries; (c) the B/W trace. Bottom row - identification of n-sided cells.

However, the irreversibility of coarsening can be twofold: changes in the froth matter (aging effects such as drying and drainage) and/or changes in the froth topology. Whilst it is impossible to reverse time, in the case of the suprafroth, however, the equivalent controlling parameter, magnetic field can be increased and decreased so as to examine the topological elasticity of the structure, while the physical properties of cell walls remain perfectly reversible. Figure 4 examines the effect of minor loop in applied field on the suprafroth. In the experiment, $H=$ 466 Oe was applied after zero field cooling to $T=4.0$ $K$ and an image taken (an underlying black trace in each of the three panels of Figure 4). Then the magnetic field was increased by the amount shown and returned back to 466 Oe and new image (shown in red) acquired. An immediate (and experimentally new compared to regular froth) conclusion is that the regions with most uniform distribution of cell sizes (shown by grey square in Figure 4) are more robust compared to the regions with broader distributions. This result is a directly related to the empirical Aboav-Weaire law[1] as applied to the dynamics of cells whose neighbours have certain number of sides. Moreover, after the initial cycle, $H \rightarrow H + \Delta H \rightarrow H$, the structure remains perfectly elastic and reproducible (for small $\Delta H \sim 5\,\text{Oe}$) upon several subsequent cycles. Similar elasticity is observed if the magnetic field was decreased by a small amount and returned back to the base field.

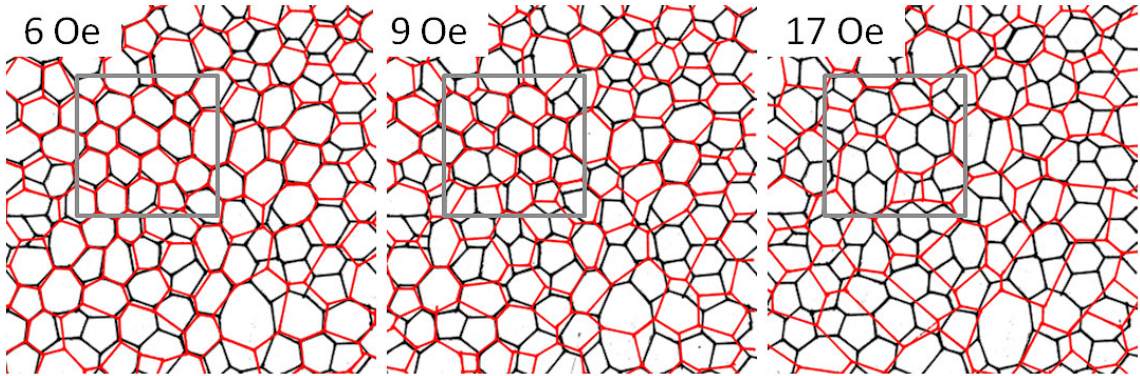

Figure 4. Structural evolution of the suprafroth at 4.0 K. Black lines correspond to the state obtained after cooling in zero field and applying 466 Oe. Red lines show the structure obtained after the field was increased by ΔH indicated in the figure and decreased back to 466 Oe. Grey square shows region of most reversible behavior.

For further comparison with conventional froths, it is important to identify the elementary transformation processes in a suprafroth. Figure 5 shows what, in physics of conventional froths, are known as T1 (side swapping) and T2 (cell disappearance) processes. These images are exactly like those observed in a coarsening soap froth. Therefore, the mesoscopic cellular dynamics in suprafroth appears to be quite similar to that of conventional froths.

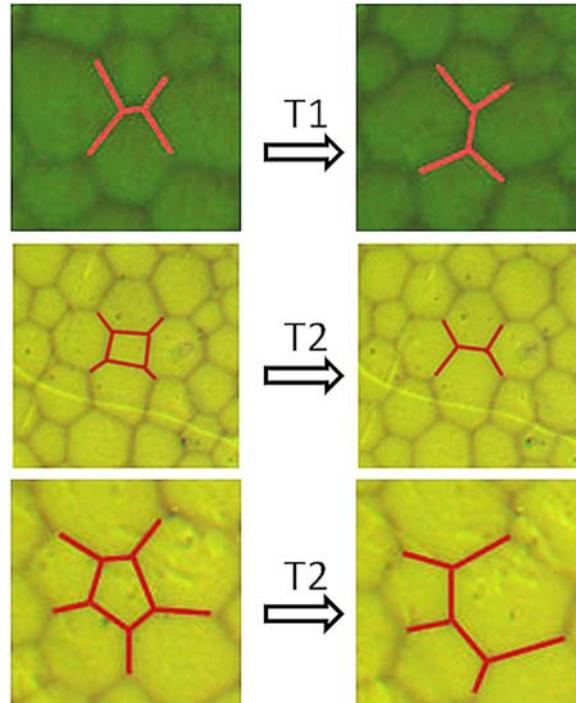

Figure 5. T1 process (top frame) measured at 4.8 K and in fields changing from 290 Oe to 300 Oe and two T2 processes imaged at 6 K in fields changing from 200 to 207 Oe.

Having examined the data in a qualitative manner, we now can proceed to a quantitative analysis of the cellular structure of the suprafroth using topological analysis similar to that depicted in Figure 3. Figure 6 shows the distribution of the number of cells with $n$-sides for suprafroth coarsened by an applied magnetic field at 4.8 K or coarsened by temperature at 320 Oe. In both cases, it is clear that $n=6$ is the most probable polygon, consistent with Euler's theorem that states that the average number of sides in a continuous two-dimensional tiling with $n$-sided cells that have 3-fold vertices is $\bar{n} = 6$ [1]. If $C$ is the total number of cells in a studied area $A$ and $C_n$ is the number of $n$-sided cells, then the distribution function $p_n = C_n/C$. As shown by a solid line, a simple triangular distribution, $p_n = 0.5(1 - 0.5|n - 6|)$ describes the observation quite well. Importantly, $p_n$ does not depend significantly on either field or temperature.

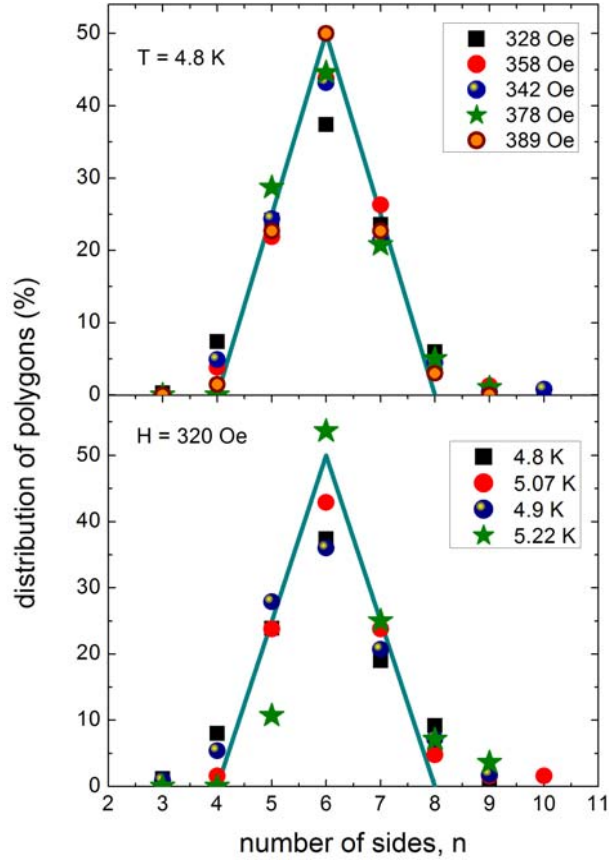

Figure 6. Distribution of the number of *n*-sided cells in froth coarsened by magnetic field (top) and temperature (bottom).

We can also examine how the average area of an *n*-sided cell depends on *n* for different magnetic fields and temperatures, Figure 7. This statistical correlation was first studied in biological, cellular structures by Lewis and is now known as Lewis's law[10],

$$A_n = \frac{A}{C} \lambda \left( n - \left[ 6 - \frac{1}{\lambda} \right] \right)$$

where the empirical constant λ is typically between 1/3 and 1, but its microscopic meaning is not well understood. It is clear from Figure 7, that in order to make our observation compatible with Lewis's law we need to set $\lambda = 1/3$, so that $A_n = (A/3C)(n-3)$.

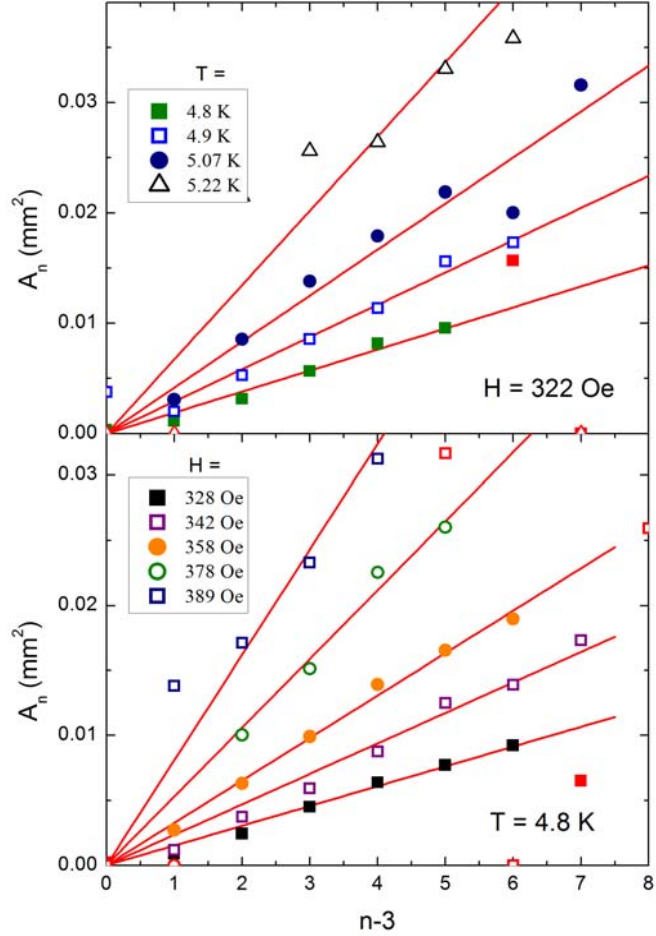

Figure 7. n-dependence of an average cell area at different temperatures (top) and fields (bottom).

Having deduced a form for the coarsening of a suprafroth, we should recall that the physics of type-I superconductor in the intermediate state dictates certain dependence of the total volume of the superconducting phase, $V_s$, on temperature and magnetic field. Specifically, in the intermediate state, the magnetic moment of a type-I superconductor of volume $V$ is $4\pi M = V\left(H - H_c\right)/N = -V_s H_c$ [11]. Therefore, the total perimeter of all cell boundaries, $P = V_s/\delta t = A\left(H_c - H\right)/\delta N H_c$ where $A$ is the total area, $t$ is sample thickness, and $\delta$ is the width of the superconducting walls in the suprafroth. It was found, by direct measurements, that $\delta \simeq 14\,\mu m$ and is practically field independent in our range of fields and temperatures. On the other hand, the total perimeter is expressed via the distribution $p_n$ and the average length of a side of an $n$-sided cell, $s_n$, as, $P = C\sum n p_n s_n/2 = 1.85\sqrt{AC}$. Therefore, we expect for the total number of cells, $C \sim \left(H_c - H\right)^2$, which is, indeed what we observe in Figure 8 (c).

Similar behaviour can be predicted for the temperature dependence of $C$, Figure 8 (d). However, the most striking result is that the $\lambda = 1/3$ coefficient of Lewis's law obtained from the linear fits shown in Figure 6 is in excellent agreement with direct, parameter-free calculation of $dA_n/dn = \lambda A/C$ with the experimental values of $A$ and $C$ as shown by solid (B-spline) line in Figure 8 (a,b). The deviations at higher temperatures are related to breaking the cellular structure apart.

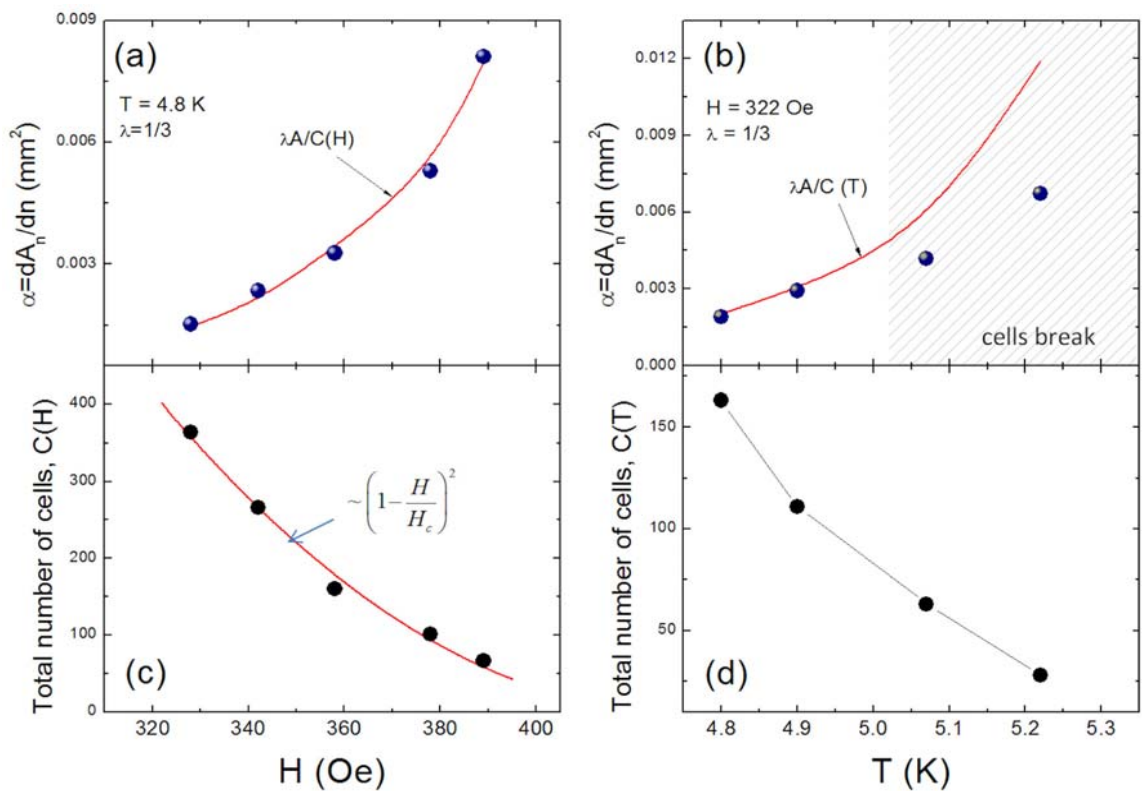

Figure 8. (top row) the coefficient in the Lewis's law. (bottom row) field and temperature dependence of the total number of cells, $C$.

One of the most studied and discussed statistical laws in the physics of froth coarsening is the von Neumann law[1,12] which predicts a linear dependence of the rate of change of the average area of an $n$-sided cell to its number of sides, $n$. In conventional froths, it has been shown both experimentally and theoretically that $dA_n/dt = \gamma (n - 6)$. Historically, the fact that the off-set to $n$ was the number six was associated with the Euler tiling theorem: six being the most probable polygon. In the case of the suprafroth we can generalize von Neumann's law and consider the temperature and magnetic field derivatives. By direct differentiation of the Lewis's law, $A_n = \left(A/3C\right)\left(n-3\right)$, with $C = A\left(H_c - H\right)^2 \big/ \left(1.85\delta N H_c\right)^2$, we expect

$$\frac{dA_n}{dH} = \frac{2.28}{H_c}\frac{\left(N\delta\right)^2}{\left(1-H/H_c\right)^3}\left(n-3\right) = \beta\left(n-3\right)$$

If the derivative is calculated for the small variation of the magnetic field (~5% in our case), so that $H/H_c \approx const$, the pre-factor $\beta$ does not vary by much. Therefore, despite that we still have hexagons as most probable polygons, the rate of change is predicted to increase for any $n>3$. This is, indeed, observed in the experiment as shown in Figure 9 where the derivatives were evaluated in the small range of fields (or temperatures).

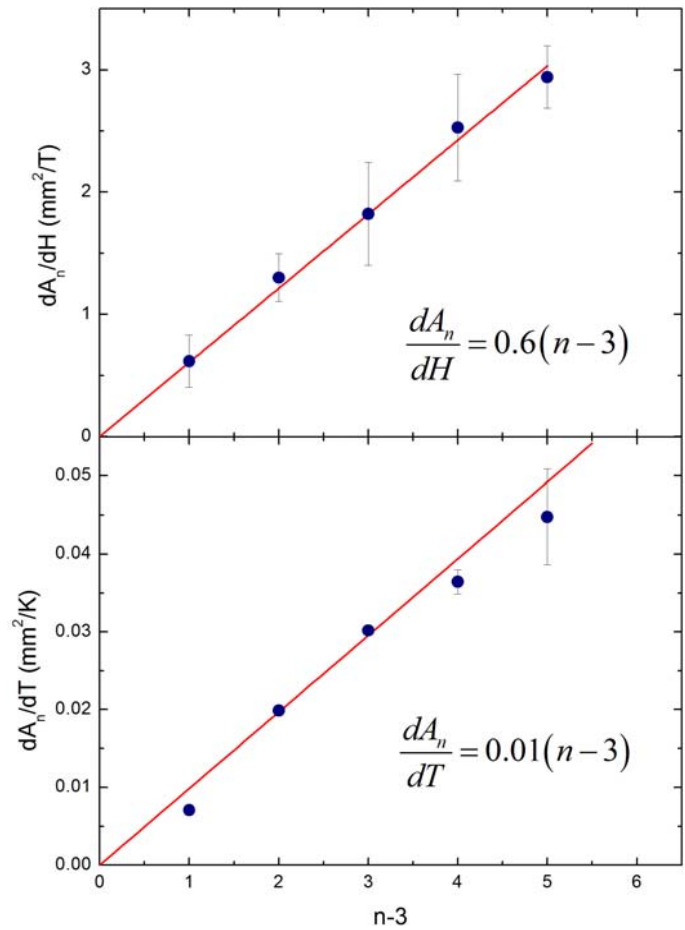

Figure 9. Generalized von Neumann's law for magnetic field and temperature derivatives. $dA_n/dH$ was calculated at $T$=4.8 $K$ and $dA_n/dT$ was calculated for $H$=330 $Oe$.

Clearly, a (*n*-3) dependence is seen in both graphs. We can even estimate the coefficient β by substituting the experimental values, $H$=0.8 $H_c$, $H_c$=0.044 $T$, δ=0.014 mm and $N$=0.63, so that we estimate $\beta \approx 0.5$ which is quite close to the observed value of 0.6 given the uncertainty in the demagnetization factor and superconducting wall width, δ (upper frame of Figure 9). It is more difficult to estimate the coefficient for $dA_n/dT$ (lower frame of Figure 9), because $M(T)$ is a more complex function of temperature.

The ultimate explanation for the observed (*n*-3) rather than (*n*-6) behavior has to be associated with the differences between the microscopic mechanisms of the coarsening. Whereas in conventional froths the amount of froth material remains constant or decreases via drainage and the process is controlled by the surface tension and diffusion of the boundaries, in suprafroth the coarsening is controlled by the magnetic field that exerts magnetic pressure inside the superconducting cells. Therefore, every polygon is inflated and the minimum planar object has $n$=3 sides, so all polygons grow. Also, the exact formula for the area of a regular polygon with side length $s_n$, $A_n = \frac{n}{4}s_n^2 \cot\frac{\pi}{n}$, can be roughly approximated by the (*n*-3) behaviour. On the other hand, the linearity of the growth rate with *n* is something that comes from general physical of the coarsening froths. To this end this work on suprafroth formally divorces the most probable number of sides (6) and the off-set in von Neumann's law (3 for the suprafroth). It should be noted that a reversed von Neumann's law due tro specifics of the interfacial interaction with the substrate was reported in ferrofluids[2,13]. This shows that dimensionality and nonlocality are important parameters determining the statistical behaviour of froths.

**Methods**

Direct visualization of the magnetic induction on sample surface was performed in a flow-type, [4]He, close-cycle cryostat using Bi-doped iron garnets with in-plane magnetization, serving as indicators[5,6]. The sample was in vacuum and placed on a copper cold stage. The indicator was placed on top of the sample. In the experiment, linearly polarized light was passed through the indicator, reflected from a mirror sputtered on its bottom to the analyzer that was turned 90 degrees with respect to the polarizer. The distribution of the magnetic moments in the indicator mimics the distribution of the magnetic induction on the surface of a studied sample, thus enabling for a direct, real-time, visualization of the magnetic flux. In all images dark corresponds to superconducting and bright to the normal phase.

Single crystal lead samples with (001) orientation were grown, cut and polished by Mateck GmbH (http://www.mateck.de/ ). The samples were in form of discs, 5 mm in diameter and 1 mm thick.

Acknowledgements: We thank John Clem, Nigel Goldenfeld, Rudolf Huebener, Vladimir Kogan, Roman Mints and Jörg Schmalian for helpful discussions. Work at the Ames Laboratory was supported by the Department of Energy, Basic Energy Sciences under Contract No. DE-AC02-07CH11358. R. P. Acknowledges a support from the NSF Grant Number DMR-05-53285 and the Alfred P. Sloan Foundation.

Correspondence and requests for materials should be addressed to Ruslan Prozorov (prozorov@ameslab.gov).